# Lattice dynamics and a magnetostructural phase transition of the nickel orthoborate Ni$_3$(BO$_3$)$_2$

R. V. Pisarev, M. A. Prosnikov, V. Yu. Davydov, A. N. Smirnov, E. M. Roginskii

*Ioffe Physical Technical Institute, Russian Academy of Sciences, 194021 St.-Petersburg, Russia*

K. N. Boldyrev, A. D. Molchanova, M. N. Popova

*Institute of Spectroscopy, Russian Academy of Sciences, 142190 Moscow, Troitsk, Russia*

M. B. Smirnov

*St.-Petersburg State University, 199034 St.-Petersburg, Russia*

V. Yu. Kazimirov

*Joint Institute for Nuclear Research, 141980 Dubna, Russia*

**Abstract**

Nickel orthoborate Ni$_3$(BO$_3$)$_2$ having a complex orthorhombic structure *Pnnm* (#58, *Z*=2) of the kotoite type is known for quite a long time as an antiferromagnetic material below $T_N$=46 K, but up to now its physical properties including the lattice dynamics have not been explored. Six [NiO$_6$] units of 2*a* and 4*f* types are linked via rigid [BO$_3$] groups and these structural particularities impose restrictions on the lattice dynamics and spin-phonon interactions. We performed the symmetry analysis of the phonon modes at the center of the Brillouin zone. The structural parameters and phonon modes were calculated using Dmol3 program. We report and analyze results of infrared and Raman studies of phonon spectra measured in all required polarizations. Most of the even and odd phonons predicted on the basis of the symmetry analysis and theoretical calculations were reliably identified in the measured spectra. A clear evidence of the spin-phonon interaction was found for some particular phonons below $T_N$. Unexpected emergence of several very narrow and weak phonon lines was observed in the infrared absorption spectra exactly at the magnetic ordering temperature $T_N$. Moreover, anomalous behavior was found for some Raman phonons. Emergence of new phonon modes in the infrared and Raman spectra exactly at $T_N$ proves existence of a magnetostructural phase transition of a new type in Ni$_3$(BO$_3$)$_2$. A possible nature of this transition is discussed.

## I. INTRODUCTION

Oxyborates of transition metals and/or rare-earth metals crystallize in a large variety of crystal structures and many of them present interesting physical properties. These materials are characterized by structural complexity due to the presence of planar [BO$_3$] or tetrahedral [BO$_4$] highly-covalent groups. Depending on particular structural details, these groups play important role in the lattice dynamics, in modifying the electronic structure, and in enhancing, decreasing or even destroying interactions between magnetic ions. Different arrangements of the boron groups result in different dimensionalities of crystallographic units and magnetic interactions and lead to a rich variety of magnetic, optical, magneto-optical, acoustical, and other properties. A good compilation of experimental results on physical properties of magnetic and non-magnetic oxyborates published before the year 1993 was given in Refs. [1,2].

Many oxyborates of transition and/or rare-earth metals have crystal structures which originate from or are closely related to minerals. In the trigonal mineral *jeremejevite* (*eremeevite*) AlBO$_3$ [3,4] the diamagnetic Al$^{3+}$ ions can be totally replaced by the magnetic Fe$^{3+}$ ions resulting in an iron borate FeBO$_3$, which becomes antiferromagnetically ordered below $T_N$=348 K [5]. Another iron borate Fe$_3$BO$_6$, which is an antiferromagnet below $T_N$=508 K, is isostructural with the mineral *norbergite* Al$_3$BO$_6$ [4,6]. Most of the other transition-metal oxyborates are antiferromagnets with transition temperatures much below the room temperature. In a magnesium-iron borate mineral *ludwigite* Mg$_2$FeBO$_5$ [4,7] the diamagnetic Mg$^{2+}$ ions can be substituted by the magnetic Mn$^{2+}$, Fe$^{2+}$, Ni$^{2+}$, and other ions making these materials magnetically ordered. Several oxyborates with similar chemical composition crystallize in the *warwickite*-type structure, for example, MgFeBO$_4$ or a mixed-valence compound Fe$_2$BO$_4$ [4]. The orthorhombic mineral *kotoite* Mg$_3$(BO$_3$)$_2$ [8,9] is another example of oxyborates in which the diamagnetic Mg$^{2+}$ ions can be substituted by several two-valence 3$d$ ions leading typically to an antiferromagnetic ordering in the range of $T_N$ ~ 10-50 K [1,2,10-12]. A very interesting particular case is a copper metaborate CuB$_2$O$_4$ known for more than a century [13] but only recently discovered as a mineral and named *santarosaite* [14]. This material crystallizes in the non-centrosymmetric tetragonal space group $I$-42$d$ [15] and demonstrates a rich variety of interesting and, in some sense, unique magnetic and optical properties (see, e.g., Refs. [16-20] and references therein). Surprisingly, only CuB$_2$O$_4$ crystallizes in this crystal structure. Chemically similar transition-metal oxyborates $M$B$_2$O$_4$ ($M$ = Mn, Fe, Co, and Ni) can be synthesized only under high-pressure and high-temperature conditions but typically they possess a monoclinic structure; (see, e.g., the recent publication on MnB$_2$O$_4$ [21] and references therein).

There are two large interesting groups of oxyborates which show multiferroic properties. In the mineral *boracite* Mg$_3$B$_7$O$_{13}$Cl [2,22], magnesium Mg$^{2+}$ ions can be replaced by bivalent 3$d$-magnetic ions resulting in numerous materials with magnetoelectric and multiferroic properties [1]. The very recent publication [23] gives a remarkable review of this group of materials. During the last decade, a large amount of studies was devoted to structural, magnetic, dielectric, and multiferroic properties of the rare-earth (*R*) oxyborates $RM_3(BO_3)_4$, where $M$ = $Fe^{3+}$, $Cr^{3+}$, $Al^{3+}$. These materials crystallize in the non-centrosymmetric structure of the mineral *huntite* $CaMg_3(CO_3)_4$ [24,25]. Other examples of complex magnetic oxyborates can be found in the literature, for instance, an actively studied quantum antiferromagnet $SrCu_2(BO_3)_2$ [26,27], $PbMBO_4$ ($M$ = $Cr^{3+}$, $Mn^{3+}$, and $Fe^{3+}$) [28,29], a recently synthesized $LiMBO_3$ ($M$ = $Mn^{2+}$, $Fe^{2+}$, $Co^{2+}$) [30], and many others.

There are several cases which show that magnetic oxyborates possess optical properties noticeably different from those of simple transition-metal oxides. Let us fix on a couple of examples. Thus, $Fe^{3+}$ ion oxides, such as $FeBO_3$ and $GdFe_3(BO_3)_4$, are highly transparent magnetic materials in the visible spectral range [31,32], whereas the iron oxide *hematite* α-$Fe_2O_3$ with the same crystal structure as $FeBO_3$ is completely opaque. On the other hand, $Fe_3BO_6$ is also opaque, similar to α-$Fe_2O_3$, because of a larger relative concentration of iron ions in comparison to $FeBO_3$. In Sec. IID, we show that the chemical "diluting" the cubic antiferromagnet NiO by [$BO_3$] groups leads to strong differences between the optical properties of NiO and $Ni_3(BO_3)_2$. Another example is an opaque multiferroic antiferromagnet CuO with the band gap of ~1.5 eV [33,34]. Broad *d-d* electronic bands due to transitions between the states of the $Cu^{2+}$ ions in the crystal field are observed in RIXS experiments [35]. In contrast, $CuB_2O_4$ which can be regarded as CuO "diluted" with [$BO_3$] groups demonstrates a unique exceptionally rich fine electronic and vibronic structure of *d-d* transitions [19,36,37]. *Huntite*-type rare-earth-iron oxyborates show interesting optical properties of 4*f* ions [38]. We conjecture that unusual optical properties of magnetic borates are, at least in part, related to a particular distribution of electronic density due to the presence of strongly covalent [$BO_3$] and [$BO_4$] groups which serve as bridges between magnetic groups. Several specific spectroscopic features of electronic 3*d* and 4*f* transitions in oxyborates open new interesting opportunities for the studies of magnetic and multiferroic materials (see, e.g., Refs. [17,18,32,38-41]).

In this paper, we present results on theoretical and experimental studies of the lattice dynamics of $Ni_3(BO_3)_2$. An antiferromagnetic (AFM) ordering in manganese, cobalt, and nickel kotoites was reported for powder samples in Ref. [12]. Only recently, the results obtained on single crystals of $Ni_3(BO_3)_2$ and $Co_3(BO_3)_2$ answered some questions about magnetic properties of these compounds [10,11]. However, many other properties of magnetic kotoites remain unexplored. Up to now, no reports are available on phonon dynamics, electronic structure, and optical properties of $Ni_3(BO_3)_2$. In our paper, we analyze the lattice dynamics at the $\Gamma = 0$ point of the Brillouin zone (BZ) using the symmetry principles. We report and analyze results on infrared reflection and absorption spectra and Raman scattering spectra. The experimental results are compared with *ab initio* theoretical calculations. We succeeded in finding all even and odd phonons at the Γ = 0 BZ point of this complex material. An unknown structural phase transition has been found at the temperature $T_N$=46 K of the antiferromagnetic transition, which evidences an intriguing intrinsic coupling between the lattice dynamics and magnetic ordering.

The paper is organized as follows. In Section II, we discuss the crystal structure of $Ni_3(BO_3)_2$ and give the symmetry analysis of the phonon modes. Section III is devoted to the description of experimental and computational details. In Section IV, we present and discuss theoretical and experimental results. Conclusions are given in Section V.

## II. CRYSTAL STRUCTURE AND SYMMETRY ANALYSIS OF THE PHONON MODES

### A. Description of the crystal structure

The nickel orthoborate $Ni_3(BO_3)_2$ and similar materials with $Ni^{2+}$ ion being replaced by the magnetic ions $Mn^{2+}$ and $Co^{2+}$ ions and non-magnetic $Mg^{2+}$ ion are known for quite a long time and their crystal structures at ambient temperature were analyzed in several publications [22,42-45]. These materials crystallize in the orthorhombic system with the *kotoite*-type structure. The point group is *mmm* ($D_{2h}$), the space group is *Pnnm* (#58), with two formula units in the unit cell. The choice of the axes and their unit cell values slightly differ in published papers. In our paper, we adopt the choice of the *Pnnm* group with the lattice parameters $a$=5.396 Å, $b$=4.459 Å, and $c$=8.297 Å, according to Ref. [42].

Four pictures of the unit cell of $Ni_3(BO_3)_2$ are shown in Fig. 1. Projections along the three crystallographic axes allow one to get more clear pictures of coordination features. The anionic structure consists of the corrugated oxygen planes perpendicular to the [100] direction. These planes include distorted triangular [$BO_3$] units [see Fig 1(b)]. Boron $B^{3+}$ ions occupy the 4$g$ positions with the *m* site symmetry. The average B-O distance is equal to 1.38 Å which is close to the value 1.37 Å typical for other borates [46]. The chemical bonding in these [$BO_3$] units is predominantly covalent.

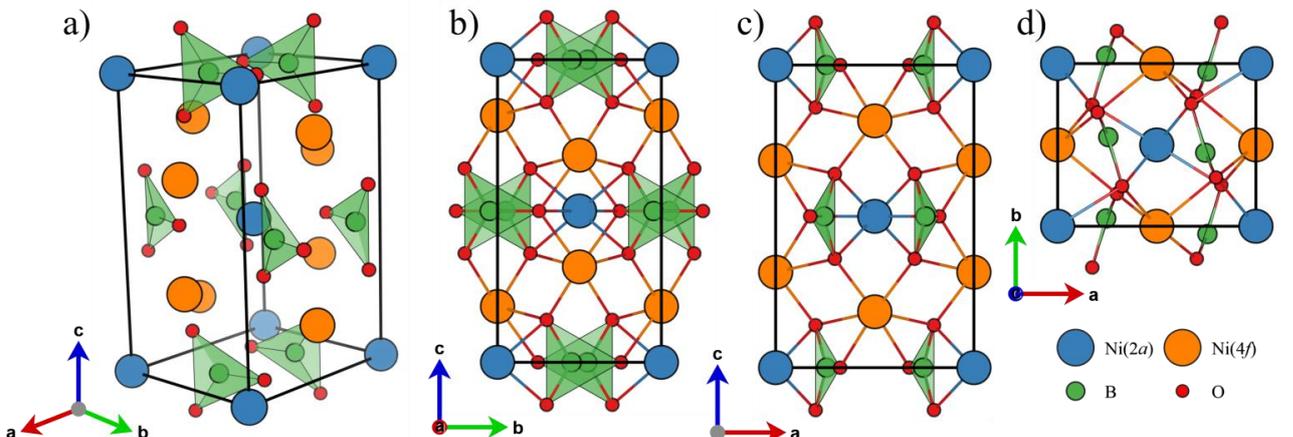

FIG. 1. (Color online) Visualizations of the $Ni_3(BO_3)_2$ unit cell. (a) Isometric view, and projections of the unit cell, (b) along the *a* axis, (c) along the *b* axis, and (d) along the *c* axis.

The Ni$^{2+}$ ions in the unit cell are situated between the anionic planes and occupy two different, Ni(2*a*) and Ni(4*f*), octahedral positions thus forming two nonequivalent sublattices. The bonding within the both [NiO$_6$] groups is predominantly ionic. We note that the 2*a* positions have the 2/*m* symmetry possessing the center of inversion, and therefore the Ni(2*a*) ions do not contribute to the Raman scattering process, whereas the 4*f* positions have the symmetry 2 of the twofold axis without the center of inversion. Octahedral arrangements at the both positions are strongly distorted. These two octahedral groups are linked together by their own edges and [BO$_3$] units forming a three-dimensional network. Therefore, Ni$_3$(BO$_3$)$_2$ can be considered as an ionic crystal and when analyzing its phonon spectra we can separate the internal [BO$_3$] modes and the external lattice modes which include rotations and translations of the [BO$_3$] units and the displacements of the Ni$^{2+}$ ions. More detailed description of the crystal structure with the bond lengths and the valence angles can be found in Refs. [42-45].

### B. Symmetry analysis of the phonon modes

Ni$_3$(BO$_3$)$_2$ belongs to the point group *mmm*, space group *Pnnm* (#58, Z=2) and in our notations the *z*(*c*) axis is a particular axis. There are 22 atoms in the primitive cell and the symmetry analysis leads to the following distribution of the 66 phonon modes between the irreducible representations at the $\Gamma = 0$ point of the Brillouin zone:

$$\Gamma = 8A_g\ (xx,yy,zz) + 8B_{1g}(xy) + 7B_{2g}(xz) + 7B_{3g}(yz) + 7A_u + 7B_{1u}(z) + 11B_{2u}(y) + 11B_{3u}(x). \quad (1)$$

After subtracting $B_{1u}(z) + B_{2u}(y) + B_{3u}(x)$ acoustic modes, one gets 63 optical phonon modes. All four types of the Raman modes, twenty six infrared polar modes, and three acoustic modes are non-degenerate. The seven $A_u$ modes are silent. All four corresponding Raman tensors are symmetric and have the following forms:

$$A_g = \begin{pmatrix} a & 0 & 0 \\ 0 & b & 0 \\ 0 & 0 & c \end{pmatrix},\ B_{1g} = \begin{pmatrix} 0 & d & 0 \\ d & 0 & 0 \\ 0 & 0 & 0 \end{pmatrix},\ B_{2g} = \begin{pmatrix} 0 & 0 & e \\ 0 & 0 & 0 \\ e & 0 & 0 \end{pmatrix},\ B_{3g} = \begin{pmatrix} 0 & 0 & 0 \\ 0 & 0 & f \\ 0 & f & 0 \end{pmatrix}. \quad (2)$$

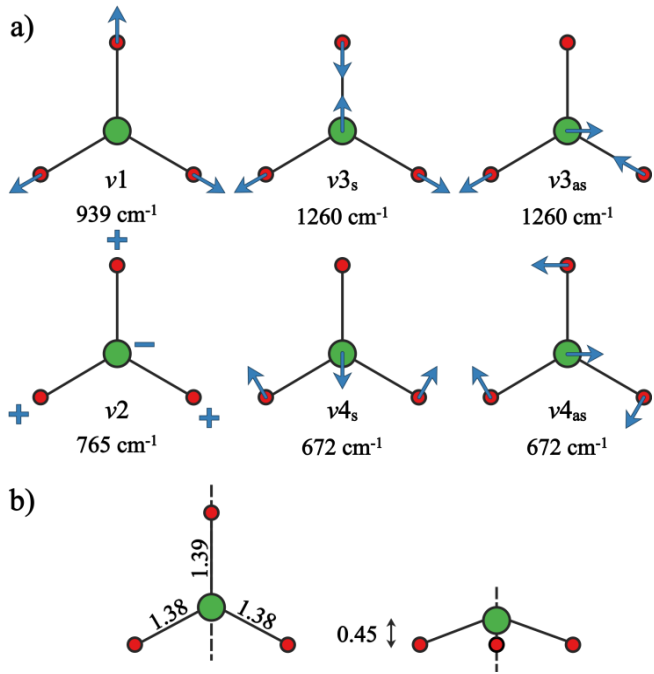

FIG. 2. (Color online) (a) Six types of the normal modes of the $[BO_3]^{3-}$ ideal planar anion with the $\bar{6}m2$ symmetry [47]. (b) Distorted $[BO_3]$ group with the $m$ symmetry (dashed line). Numbers mark bond lengths (Å), and out-of-plane shift of the boron ion.

The $[BO_3]$ units have the $m$ symmetry. However, they are only slightly distorted with respect to the $\bar{6}m2$ planar configuration. Hence, one can suggest that their internal vibrations must roughly obey the symmetry selection scheme proposed for an isolated $[BO_3]^{3-}$ anion (see Fig. 2). Frequencies of normal vibrations of this anion were listed in Ref. [47]. In the crystal lattice of $Ni_3(BO_3)_2$, there are four $[BO_3]^{3-}$ anions in the unit cell, which results in the Davydov (factor-group) splitting of each free-molecule vibration into four crystalline modes. The correlational analysis shows that the $\nu1$ and $\nu2$ modes, which are symmetric in respect to the $m$ plane, should manifest themselves in the $A_g$, $B_{1g}$, $B_{2u}$, and $B_{3u}$ representations at about 939 and 765 cm$^{-1}$, respectively. The doubly degenerate $\nu3$ and $\nu4$ modes should split into symmetric and antisymmetric components and appear at about 1260 and 672 cm$^{-1}$, respectively. The symmetric components are expected in the $A_g$, $B_{1g}$, $B_{2u,}$ and $B_{3u}$ representations, whereas the antisymmetric components must contribute to the $B_{2g}$, $B_{3g}$, $A_u$, and $B_{1u}$ representations. Thus, each Raman-active vibration of the free $[BO_3]$ group generates a Raman doublet in the crystal, either $(A_g+B_{1g})$ or $(B_{2g}+B_{3g})$, but each IR-active $[BO_3]$ vibration transforms into either a doublet $(B_{2u}+B_{3u})$ or a singlet $B_{1u}$ in the infrared spectrum of the crystal. All modes of the $[BO_3]$ group, derived using the correlation analysis, are summarized in Fig. 3.

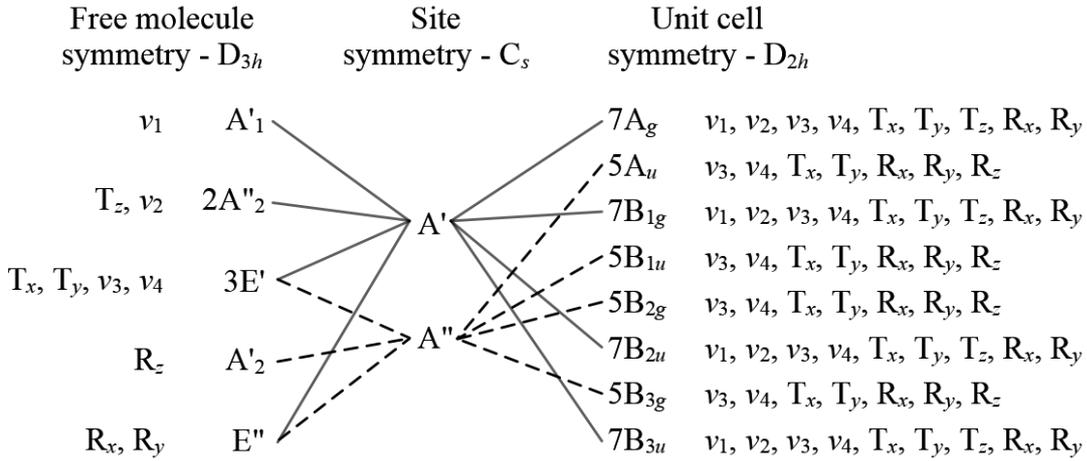

FIG. 3. Correlation scheme for the [BO$_3$] group placed into the C$_s$ symmetry position of the Ni$_3$(BO$_3$)$_2$ unit cell having the D$_{2h}$ symmetry.

Among twelve Ni(2$a$)-O bond-stretching modes, one can discriminate three pairs of transverse optical TO-like modes, namely, $2(B_{2u}+B_{3u}) + (A_u+B_{1u})$, and six other modes $(2A_g+B_{2g}+2B_{1g}+B_{3g})$, in which Ni(2$a$) atoms do not oscillate. The Ni-O bond lengths in our compound are close to those in the cubic NiO crystal (2.089 Å) [48]. Hence, one can suppose that the TO-like Ni-O bond-stretching modes have frequencies of about 400 cm$^{-1}$, close to the TO frequency of the NiO crystal [49]. The twenty four Ni(4$f$)-O bond-stretching modes are regularly distributed over all eight symmetry representations with three modes in each representation. No symmetry-based selection rules exist for the Ni(4$f$)-O TO-like modes. In our analysis of the results we will assume that all phonon modes with frequencies below ~ 400 cm$^{-1}$ are external lattice modes. This means that within these modes the [BO$_3$]$^{3-}$ anions oscillate as quasi-rigid units and we should analyze rotations $R$ and translations $T$ of the [BO$_3$] groups and the displacements of the Ni$^{2+}$ ions (see Tables II-VIII).

### III. EXPERIMENTAL AND COMPUTATIONAL DETAILS
#### A. Infrared reflection and transmission experiments

Infrared reflection and transmission spectra were registered in a broad spectral range using a Fourier-transform IR spectrometer Bruker IFS 125HR. The measurements were performed for different temperatures between 3 and 300 K using a closed-cycle helium cryostat Cryomech ST403. Helium-cooled bolometer for the far infrared (FIR) spectral region 10 – 500 cm$^{-1}$ and a liquid-nitrogen-cooled MCT detector for the middle infrared (MIR) spectral range 400 – 3000 cm$^{-1}$ were used. A wire-grid and a KRS-5 polarizers were used in the FIR and MIR spectral regions, respectively.

Modeling of the reflection spectra using the RefFIT programs [50] allowed us to find parameters of the phonons. The spectra were fitted by the least-squares method and calculations were performed according to the equation:

$$R(\omega) = \left|\frac{\sqrt{\varepsilon(\omega)}-1}{\sqrt{\varepsilon(\omega)}+1}\right|^2, \qquad (4)$$

where $R(\omega)$ is the reflection coefficient, $\varepsilon(\omega)$ is the dielectric function, represented in the form of a sum of independent damped oscillators as

$$\varepsilon(\omega) = \varepsilon_\infty + \sum_{j=1}^{N} \frac{f_j \omega_j^2}{\omega_j^2 - \omega^2 + i\gamma_j \omega}. \qquad (5)$$

Here $N$ is the total number of the oscillators, $\omega_j$, $f_j$, and $\gamma_j$ are the frequency, the oscillator strength, and the damping constant of the *j*-th oscillator.

### B. Raman scattering experiments

Raman scattering spectra were measured in the range 40 – 1600 cm$^{-1}$ with the use of a Jobin-Yvon T64000 spectrometer equipped with a cooled CCD camera. The argon-laser line 514.5 nm (2.41 eV) and the Nd:YAG-laser line 532 nm (2.33 eV) were used for the excitation. A 50× objective was employed both to focus the incident beam and to collect the scattered light. Low-temperature spectra were recorded using a helium closed-cycle cryostat (Cryo Industries, Inc.). All measurements were done in the back-scattering geometry for various polarization settings.

### C. Ab initio density functional theory (DFT) calculations

In view of a mixed covalent-ionic bonding inherent to the Ni$_3$(BO$_3$)$_2$ compound we decided to use a computational scheme based on the atomic-like basis sets. This implies using the DMol3-program [51] and the Perdew-Wang exchange-correlation functional [52] within the local-density approximation. The double-numerical with polarization (DNP) atomic basis [53] was applied to ensure high accuracy of the computations. The DFT semi-core pseudo-potential method was chosen as a compromise between computational cost and accuracy of the simulations. The Monkhorst–Pack grid [54] of 3×3×2 points in the *k*-space was used for the Brillouin zone sampling. In fact, we have tested other DFT functionals and other basis sets implemented in various program packages (CASTEP, ABINIT, CRYSTAL). Finally, we have chosen the presented below DMol3 results as best agreeing with the available experimental data

### D. Samples

Several compounds are formed in the NiO-$B_2O_3$ system. The nickel borate $Ni_3(BO_3)_2$ is known for more than a century [13], whereas other borates, $NiB_2O_4$ and $NiB_4O_7$, were synthesized only recently, under high-pressure and high-temperature conditions [55,56]. For our studies, single crystals of $Ni_3(BO_3)_2$ were grown by a gas-transport method in the evacuated quartz ampoules along with the nickel boracites single crystals [16]. A polished rectangular-shaped sample with dimensions of about 2×4×1, mm with the main orientations along the *a(x)*, *b(y)*, and *c(z)* axes was used for optical characterization and both Raman scattering and infrared reflection and absorption measurements.

For a preliminary optical characterization of this sample, we measured the ellipsometric functions ψ and Δ in the frequency range from 0.6 to 5.6 eV. To make a comparison, similar measurements were performed for a cubic NiO(100) single crystal. The calculated dielectric functions $\varepsilon_1$ and $\varepsilon_2$ and indices of refraction *n* and absorption *k* for both materials are presented in Fig. 4. The data show a drastic drop of all these parameters in $Ni_3(BO_3)_2$ in comparison with NiO. On the other hand, the well-established fundamental band gap 4 eV in NiO [57,58] is markedly shifted in $Ni_3(BO_3)_2$ (to a value more than 5 eV, see Fig. 4). Evidently, these changes are due to a lower Ni concentration in the borate. We also have performed preliminary measurements of the optical absorption. Presence of three strong absorption bands due to the *d-d* transitions within the $Ni^{2+}$ $3d^8$ states in the crystal field, from the ground $^3A_2$ state to the $^3T_2$ (~1.07 eV), $^3T_1^a$ (1.73 eV), and $^3T_1^b$ (3.05 eV) excited states was established [59,60]. These bands and a "green" transparency window around 2.3 eV are similar to respective features observed in NiO and in other nickel oxides [59-61]. The transparency window defines dark-green color of NiO and $Ni_3(BO_3)_2$. We note that these electric-dipole forbidden transitions are practically unobservable in the spectroscopic ellipsometric measurements, as it is seen in Fig. 4. Diffuse reflection spectrum of a powder sample of $Ni_3(BO_3)_2$, with spectral features due to the *d-d* transitions, similar to those observed in our absorption measurements was published in Ref. [62]. We add that the above mentioned high-pressure borate $NiB_2O_4$ shows similar optical features and is characterized as "light-green" [55].

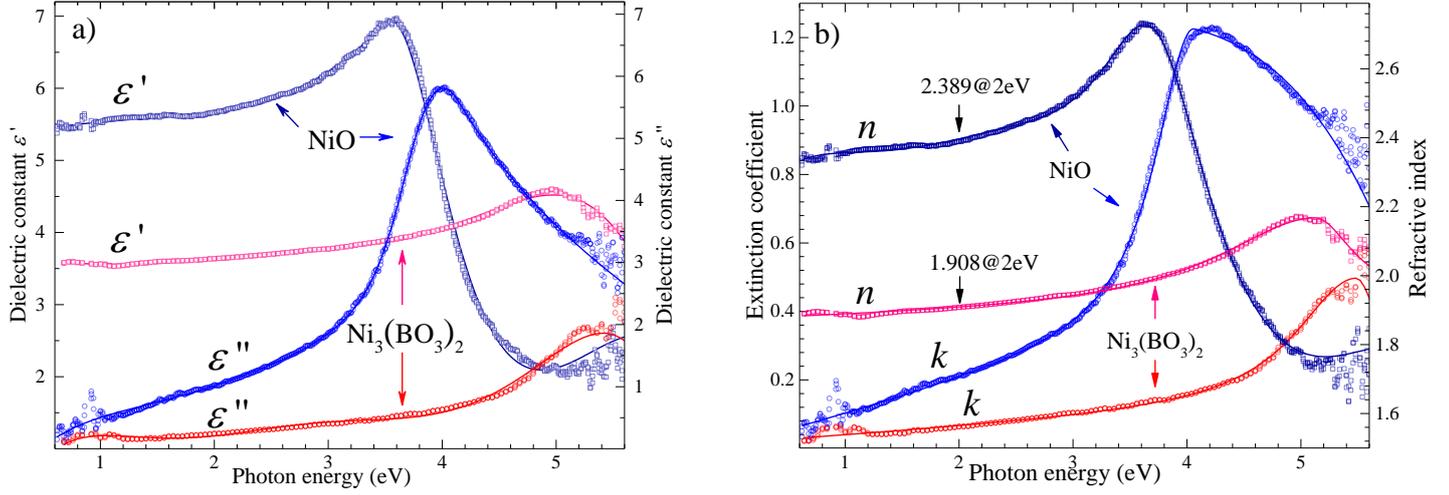

FIG. 4. (Color online) Results of the spectroscopic ellipsometry measurements for E∥$x(a)$ of the orthorhombic Ni$_3$(BO$_3$)$_2$ and cubic NiO. (a) Dielectric functions $\varepsilon_1$ and $\varepsilon_2$. (b) Indices of refraction $n$ and absorption $k$. Lines are guides to the eye.

## IV. RESULTS AND DISCUSSION

### A. DFT modeling of the structure and phonon states

The first stage of the calculations dealt with the geometry optimization, the obtained structure parameters are listed in Table 1. An overall agreement is quite good, though the computational results slightly underestimate the cell parameters (by ~ 0.1%).

TABLE I. Calculated optimized unit-cell parameters (Å), cell volume (Å$^3$), and atomic positions in comparison with the experimental data [42].

|          | DFT    |        |        | Experiment [42] |        |        |
|----------|--------|--------|--------|--------|--------|--------|
| $a$      | 5.39419 |       |        | 5.396 |        |        |
| $b$      | 4.45751 |       |        | 4.459 |        |        |
| $c$      | 8.29422 |       |        | 8.297 |        |        |
| Volume   | 199.43 |        |        | 199.63 |       |        |
| Ni (2$a$) | 0      | 0      | 0      | 0      | 0      | 0      |
| Ni (4$f$) | 0      | 0      | 0.3150 | 0      | 0      | 0.3157 |
| B (4$g$)  | 0.2545 | 0.5418 | 0      | 0.2551 | 0.5439 | 0      |
| O1 (4$g$) | 0.3239 | 0.2440 | 0      | 0.3243 | 0.2489 | 0      |
| O2 (8$h$) | 0.2018 | 0.7023 | 0.1403 | 0.2011 | 0.7012 | 0.1399 |

The zone-center phonon states were calculated at the theoretical equilibrium geometry. All frequencies were found to be real. This result confirms stability of the lattice; see, however, Sec. IVD where we discuss a magnetostructural phase transition at $T_N$. The calculated frequencies and symmetry assignments of the IR and Raman phonons are listed in Tables II-VIII. References to the corresponding [BO$_3$] internal modes are shown in the columns "Assignment" of Tables. It is seen that the frequency and symmetry distributions of these modes obey well the predictions put forward in Sec. III, even though the frequency intervals between the $\nu_2$ and $\nu_4$ groups are very close.

Atomistic pattern of the lattice modes below 400 cm$^{-1}$ is shown in Tables II-VIII by indicating the predominant types of motions, namely, translations $T_x$, $T_y$, and $T_z$ of the Ni ions and [BO$_3$] units and rotations $R_x$, $R_y$, and $R_z$ of the latter. One can see that the lowest-frequency modes involve predominantly translations, whereas the highest-frequency lattice modes correspond to librations of the [BO$_3$] units. The modes with antiparallel translations of anions and cations, which correspond to the largest oscillations of the dipole moment, are highlighted in bold. Hence, the assignments of the observed spectral lines to the calculated phonon modes were done relying on the frequency and symmetry correspondence. The established relations are discussed below.

## B. Infrared reflection spectra

Figure 5 shows the infrared reflection spectra of Ni$_3$(BO$_3$)$_2$ measured at room temperature for the three main polarizations of light, E($\omega$) ∥ $x$, $y$, and $z$. All the three spectra are different, thus demonstrating a distinct infrared anisotropy as a consequence of the orthorhombic crystal symmetry. The modeling of the reflection spectra, performed as described in Section IIIA resulted in the three sets of phonon parameters (TO frequencies $\omega_j$, damping coefficients $\gamma_j$, and oscillator strengths $f_j$) for the IR-active phonons of the three different symmetries, $B_{1u}$, $B_{2u}$, and $B_{3u}$, as well as in the values of optical high-frequency dielectric constants $\varepsilon_{\infty i}$, $i=x$, $y$, and $z$, for the three polarizations of the light E($\omega$) ∥ $x$, $y$, and $z$, respectively. Using thus obtained sets of $f_j$, we have calculated the static dielectric constants $\varepsilon_{0i}$, according to Eq. (5). Tables II-IV list the phonon parameters and dielectric constants obtained from the experimental reflection spectra. The data on phonon frequencies are compared with the results of theoretical calculations.

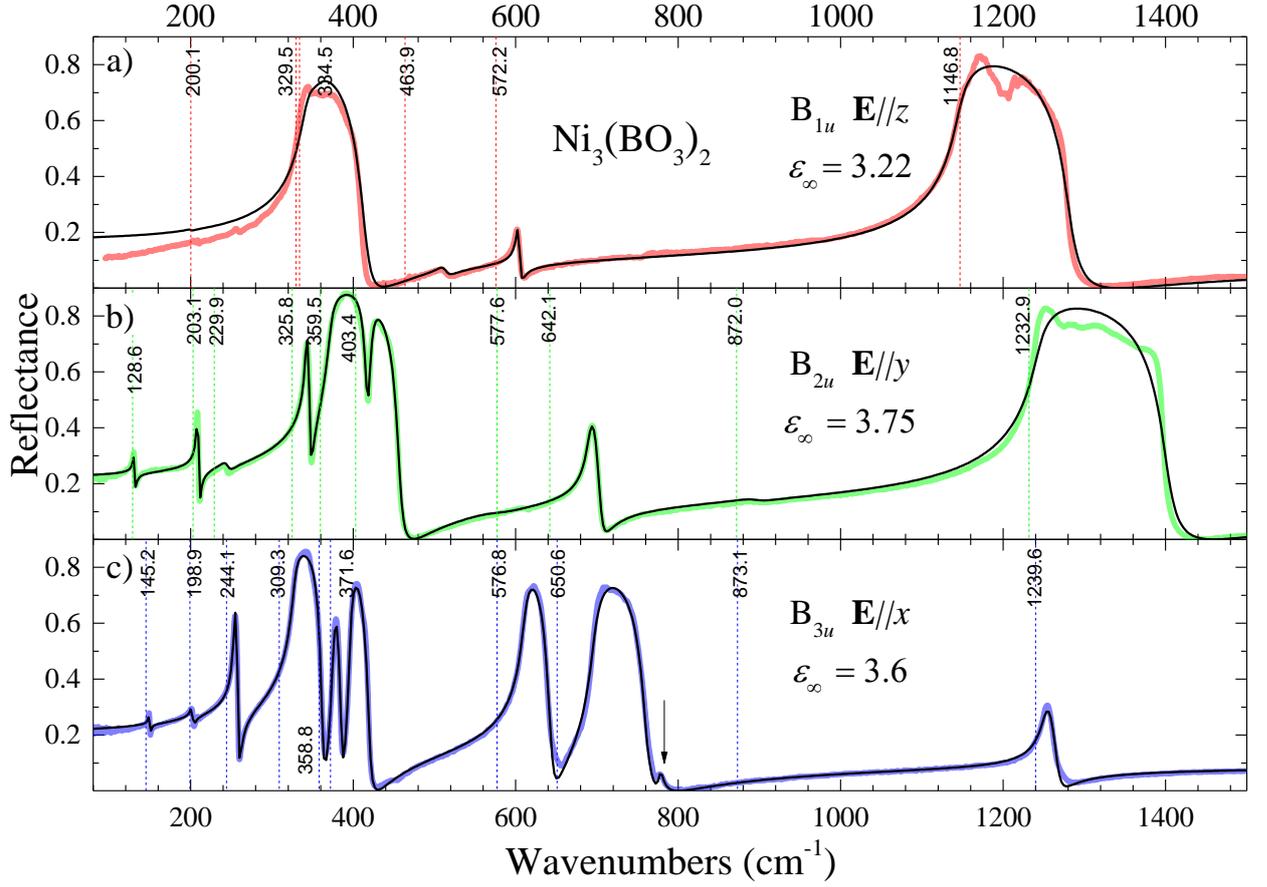

FIG. 5. (Color online) Room-temperature infrared reflection spectra of $Ni_3(BO_3)_2$ measured with the polarizations of the incident light (a) $E(\omega)\|z$ (thick red line), which probes $B_{1u}$ modes, (b) $E(\omega)\|y$ (thick green line), which probes $B_{2u}$ modes, and (c) $E(\omega)\|x$ (thick blue line), which probes $B_{3u}$ modes. Here, an arrow points to a feature at 777 cm$^{-1}$ (see the text, Sec. IVC). Spectra are compared with fitting calculations (thin black lines) based on a model of damped oscillators, see Eq. (5). Vertical dashed lines show frequencies obtained from the *ab initio* DFT calculations.

TABLE II. Parameters of the $B_{1u}$ modes found from the IR spectra ($T$ = 293 K) and *ab initio* DFT calculations. Frequencies $\omega_j$ (cm$^{-1}$), damping constants $\gamma_j$ (cm$^{-1}$), oscillator strengths $f_j$, and $\varepsilon_{\infty z}$ were found from IR spectra using the model of attenuated oscillators according to Eq. (4). $\varepsilon_{\infty z} = 3.22$, $\varepsilon_{0z} = 6.1$ (see text).

| DFT | Experiment | | | Assignment | | |
|---|---|---|---|---|---|---|
| $\omega_j$(TO) | | $\gamma_j$ | $f_j$ | Ni(2a) | Ni(4f) | [BO$_3$] |
| 200.1 | 200 [a] | 5.2 | 0.006 | $T_z$ | | $T_z$ |
| **329.5** | **340** | 16.4 | **1.98** | $T_z$ | $T_z$ | $T_z$ |
| 334.5 | 354 | 19.1 | 0.08 | | | $R_y$ |
| 463.9 | 513[a] | 13.6 | 0.02 | | | $R_y$ |
| 572.2 | 603 | 4.7 | 0.033 | | | $\nu 4$ |
| 1146.8 | 1146 | 20.6 | 0.765 | | | $\nu 3$ |

[a] Very weak line

TABLE III. Parameters of the $B_{2u}$ modes found from the IR spectra ($T$ = 293 K) and *ab initio* DFT calculations. $\varepsilon_{\infty y} = 3.75$, $\varepsilon_{0y} = 7.9$.

| DFT | Experiment | Assignment |
|---|---|---|

| $\omega_j$(TO) | $\gamma_j$ | $f_j$ | Ni(2a) | Ni(4f) | [BO$_3$] |
|---|---|---|---|---|---|
| 128.6 | 132 | 1.8 | 0.071 | $T_x$ | $T_x$ | $T_x$ |
| 203.1 | 208 | 2 | 0.161 | $T_y$ | $T_y$ | |
| 229.9 | 245 | 11.2 | 0.07 | | $T_x$ | $T_x$ |
| 325.8 | 342 | 3.1 | 0.65 | $T_x$ | | $T_x$ |
| **359.5** | **372** | **8.6** | **2.06** | $T_y$ | $T_y$ | $T_y$ |
| 403.4 | 420 | 7.5 | 0.054 | | | $R_z$ |
| 577.6 | 575[a] | 55.6 | 0.023 | | | $\nu_4$ |
| 642.1 | 691 | 10.5 | 0.164 | | | $\nu_2$ |
| 872.0 | 896[a] | 33.4 | 0.011 | | | $\nu_1$ |
| 1232.9 | 1244 | 20.9 | 0.938 | | | $\nu_3$ |

[a] Very weak line

TABLE IV. Parameters of the $B_{3u}$ modes found from the IR spectra ($T = 293$ K) and *ab initio* DFT calculations. $\varepsilon_{\infty x} = 3.6$, $\varepsilon_{0x} = 7.5$.

| DFT | Experiment | | | Assignment | | |
|---|---|---|---|---|---|---|
| $\omega_j$(TO) | | $\gamma_j$ | $f_j$ | Ni(2a) | Ni(4f) | [BO$_3$] |
| 145.2 | 149 | 1.0 | 0.022 | $T_y$ | | $T_y$ |
| 198.9 | 202 | 2.6 | 0.043 | $T_y$ | $T_y$ | |
| 244.1 | 255 | 2.9 | 0.443 | $T_x$ | | $T_x$ |
| **309.3** | **327** | **7.4** | **1.98** | $T_x$ | $T_x$ | $T_x$ |
| 358.8 | 375 | 5.3 | 0.187 | $T_y$ | $T_y$ | $T_y$ |
| 371.6 | 396 | 5.1 | 0.174 | | | $R_z$ |
| 576.8 | 611 | 9.9 | 0.614 | | | $\nu_4$ |
| 650.6 | 696 | 13.7 | 0.395 | | | $\nu_2$ |
| 873.1 | - | - | - | | | $\nu_1$ |
| 1239.6 | 1252 | 15 | 0.074 | | | $\nu_3$ |

Results presented in Tables II-IV show that for every computed odd mode there is a counterpart among the observed IR-active phonon modes. Some modes are very weak in the reflection spectra but are clearly visible in the absorption spectra (see, e.g. Fig.8). Frequency differences between the experimental and calculated values are ~ 17 cm$^{-1}$ on average. In each of the $B_{1u}$, $B_{2u}$, and $B_{3u}$ spectra, one very intense IR line dominates in the region of external lattice vibrations (these lines are highlighted by bold in Tables II-IV). Evidently, the corresponding phonon modes involve antiparallel oscillations of the anions [BO$_3$]$^{3-}$ and cations Ni$^{2+}$ in the three orthogonal directions, *x*, *y*, and *z*. This suggestion is confirmed by the analysis of the corresponding eigenvectors. We remind that the frequency of the TO mode in NiO is equal to ~400 cm$^{-1}$. A lower value of the corresponding frequency in Ni$_3$(BO$_3$)$_2$ may be caused by a larger mass of the anions. The value of $\varepsilon_{\infty x}$=3.6 obtained from IR experiments agrees well with the results of the spectroscopic ellipsometry measurements, see Fig. 4(a). An interesting feature is observed in the E($\omega$)||z reflection spectrum of the $B_{1u}$ modes: the highest-frequency strong reflection band originating from the $\nu_3$ internal vibration of the [BO$_3$]$^{3-}$ group is split into two equally intense bands. Most probably, we are dealing with the Fermi resonance between the vibrational mode $\nu_3$ (1146 cm$^{-1}$, $B_{1u}$) and the two-particle excitation $\nu_4$ (611 cm$^{-1}$, $B_{3u}$) + $\nu_4$ (553 cm$^{-1}$, $B_{2g}$) of the same symmetry ($B_{3u} \times B_{2g} = B_{1u}$) [63, 64] (for the frequencies of the $B_{2g}$ Raman modes, see Table VII in Sec. IVC).

### C. Raman scattering spectra

To the best of our knowledge, only one work dedicated to Raman scattering of the kotoite $M_3(BO_3)_2$ family was published [9]. The authors studied infrared and Raman unpolarized spectra of several samples of the kotoite natural mineral $Mg_3(BO_3)_2$. Both types of the spectra were very complicated, first of all, as a result of overlap of modes of different symmetries, and to some degree due to the adsorbed water and impurities inevitably present in minerals. No symmetry analysis was done. According to our analysis for $Ni_3(BO_3)_2$ [see Eq. (1)] there are thirty Raman-active modes at the Γ point, namely, eight $A_g$, eight $B_{1g}$, seven $B_{2g}$ and seven $B_{3g}$. Based on the crystal symmetry, the $A_g$ modes are expected to be observed in the diagonal polarization spectra, while $B_g$ modes should manifest themselves in the off-diagonal spectra. According to orthorhombic Raman tensors [see Eq. (2)], the intensities of the $A_g$ modes for each parallel polarization ($xx$), ($yy$) and ($zz$) should be different. We remind that no modes are expected in the scattering spectra related to the Ni(2a) ions in the centrosymmetric positions.

Room-temperature polarized Raman spectra of $Ni_3(BO_3)_2$ registered in various polarization settings are collected in Figs. 6 and 7. Results of the *ab initio* calculations are shown by vertical dashed lines with relevant frequency values written at the upper part of figures. Frequencies of the observed modes are also indicated at the spectra. We note that all Raman spectra show more lines than expected and we relate this fact to the specificity of the micro-Raman technique and to a possible small misalignment of the crystal axes. It is obvious, that results of the calculations underestimate frequencies of all Raman active phonon modes, with the mean and the largest values of discrepancy 20 and 70 cm$^{-1}$, respectively. Nevertheless unambiguous assignment of the observed modes could be done, which was the main goal of these first calculations on a complicated kotoite structure. In fact, we have tested other DFT functionals and other basis sets implemented in various program packages (CASTEP, ABINIT, CRYSTAL). Finally, we have chosen the presented DMol3 results as the best agreeing with the available experimental data. The *ab initio* and experimental results are confronted in Tables V-VIII. The most intense line for each polarization is marked in bold.

Figure 6 shows Raman spectra measured in the $y(xx)\bar{y}$, $z(yy)\bar{z}$, and $x(zz)\bar{x}$ parallel polarization settings in the spectral range from 120 to 1300 cm$^{-1}$. First two spectra in Fig. 6 are similar in their general features except small intensity differences. They are characterized, first of all, by the most intense 912 cm$^{-1}$ line, and by five weaker lines 278, 351, 403, 681, and 1238 cm$^{-1}$. Taking into account the results of the *ab initio* calculations, the line 766 cm$^{-1}$ from the doublet 766/777 cm$^{-1}$ should be assigned to one of the $A_g$ modes.

The third $x(zz)\bar{x}$ spectrum in Fig. 6(c) shows significant changes in comparison with the previous ones. First of all, an intense 238 cm$^{-1}$ line appears which is practically missing in the previous spectra. Intensity of the 402 cm$^{-1}$ line strongly increased, while intensity of the 351 and 1238 cm$^{-1}$ lines, on the contrary, decreased. Several weak narrow lines are attributed to the leakage from off-diagonal modes. Weak broad features in the region of about 600 cm$^{-1}$ and ~1500 cm$^{-1}$ can be related to multi-phonon scattering processes.

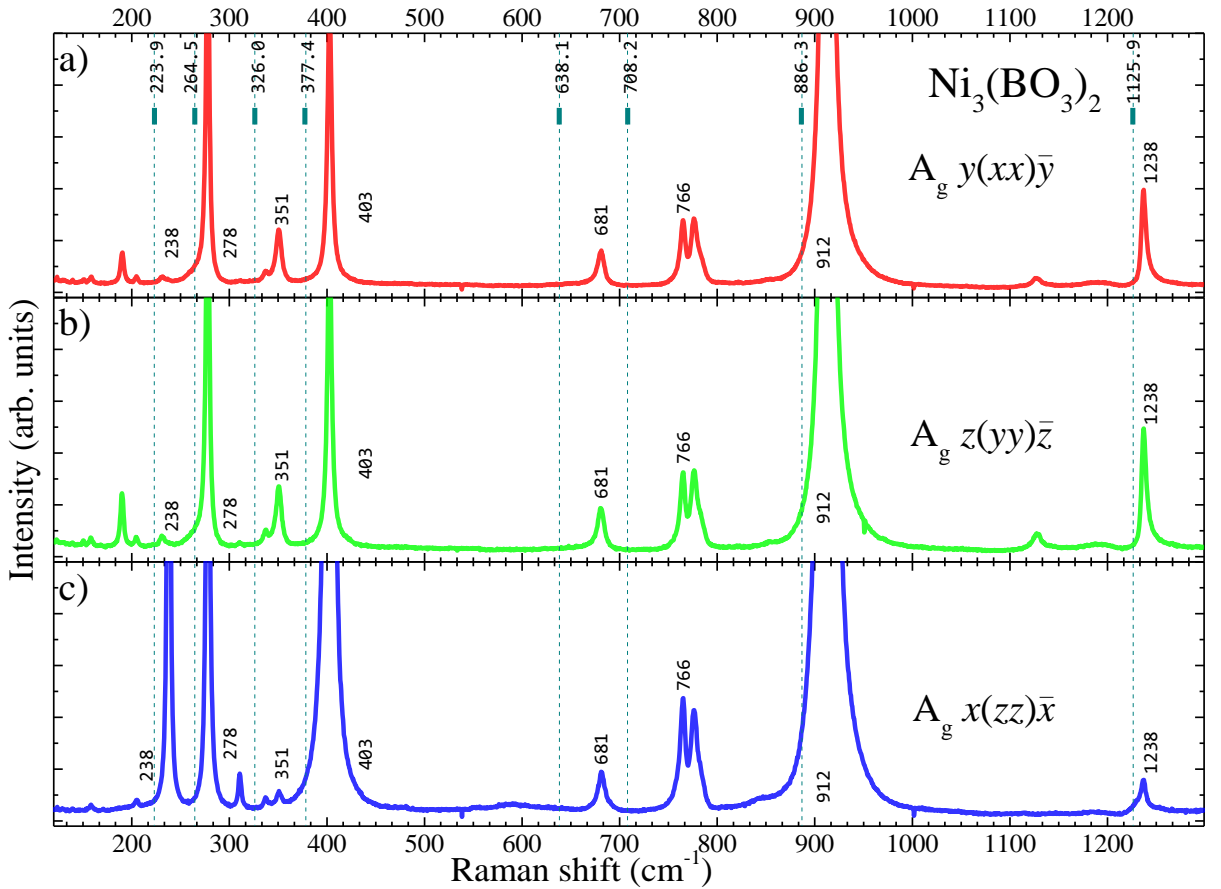

FIG. 6. (Color online) Room-temperature Raman scattering spectra of Ni$_3$(BO$_3$)$_2$ measured in parallel polarizations representing the eight $A_g$ modes measured in (a) $y(xx)\bar{y}$ (red line), (b) $z(yy)\bar{z}$ (green line), (c) $x(zz)\bar{x}$ (blue line) polarizations. Vertical dashed lines show frequencies obtained from the *ab initio* DFT calculations.

Figure 7 shows the off-diagonal Raman spectra which should reflect the eight $B_{1g}(xy)$, seven $B_{2g}(zx)$, and seven $B_{3g}(yz)$ modes. Characteristic marks of the $z(xy)\bar{z}$ spectrum are the most intense line at 310 cm$^{-1}$ and a weaker one at 690 cm$^{-1}$. We assign the high-frequency line from the 766/777 cm$^{-1}$ doublet to the $B_{1g}$ mode because the intensity of the 777 cm$^{-1}$ line exceeds that of 766 cm$^{-1}$ for the given polarization. Moreover, existence of two close modes, at 708.2 ($A_g$) and 708.7 cm$^{-1}$ ($B_{1g}$), corresponding to the Raman Davydov doublet originating from the ν2 vibration of a free BO$_3$ molecule, is predicted by *ab initio* calculations. According to calculations, a pair of closely located $A_g$ and $B_{1g}$ modes is expected at 886 and 889 cm$^{-1}$, respectively. But in the experimental spectra we observe a strong feature at 912-915 cm$^{-1}$ which should be assigned to the calculated modes. In fact, deconvolution of this feature confirms a presence of the two overlapping modes, at 912 and 915 cm$^{-1}$. This strong line can be observed in all polarizations, which may point to a partial breaking of the selection rules. In particular, a local breaking of the inversion symmetry (due to, e.g., lattice defects) could explain the presence of the mode 777 cm$^{-1}$, assigned to the $B_{1g}$ Raman-active vibration, in the $B_{3u}$ IR spectrum (see Fig. 5). Wide 1256 cm$^{-1}$ line is the highest-frequency line in the both IR and Raman spectra.

Figure 7(b) shows a Raman spectrum taken in the $y(zx)\bar{y}$ setting. There are three intense lines, at 158, 205, and 337 cm$^{-1}$, the latter being the most intense one for the given polarization. Deconvolution of the spectra has revealed two additional very weak shoulder-like lines, at 287 and 400 cm$^{-1}$. Fig. 7(c) displays a Raman spectrum in the $x(yz)\bar{x}$ setting. A characteristic mark of this spectrum is a very intense line at 189 cm$^{-1}$ with two other intense lines, at 231 and 1128 cm$^{-1}$. Also, a weak shoulder-like line at 311 cm$^{-1}$, and a broad low-intensity feature at 584 cm$^{-1}$ are present in the $B_{3g}$ spectrum. The line at 151 cm$^{-1}$ is the lowest-frequency Raman line in the spectra of Ni$_3$(BO$_3$)$_2$.

Off-diagonal polarization spectra show a leakage of intense lines from the diagonal polarization. For more reliable separation of all modes and for getting an additional information about their temperature dependence (see also Sec. IVD) we have performed scattering experiments below the room temperature. As the temperature is lowered, the phonon modes sharpen and the presence of the $B_{2g}$ lines at 287 and 400 cm$^{-1}$, and $B_{3g}$ lines at 311 and 584 cm$^{-1}$ becomes evident in the spectra. Most phonon lines show usual hardening and narrowing upon decreasing the temperature down to 10 K, however some of them behave anomalously and are discussed in the next Sec. IVD.

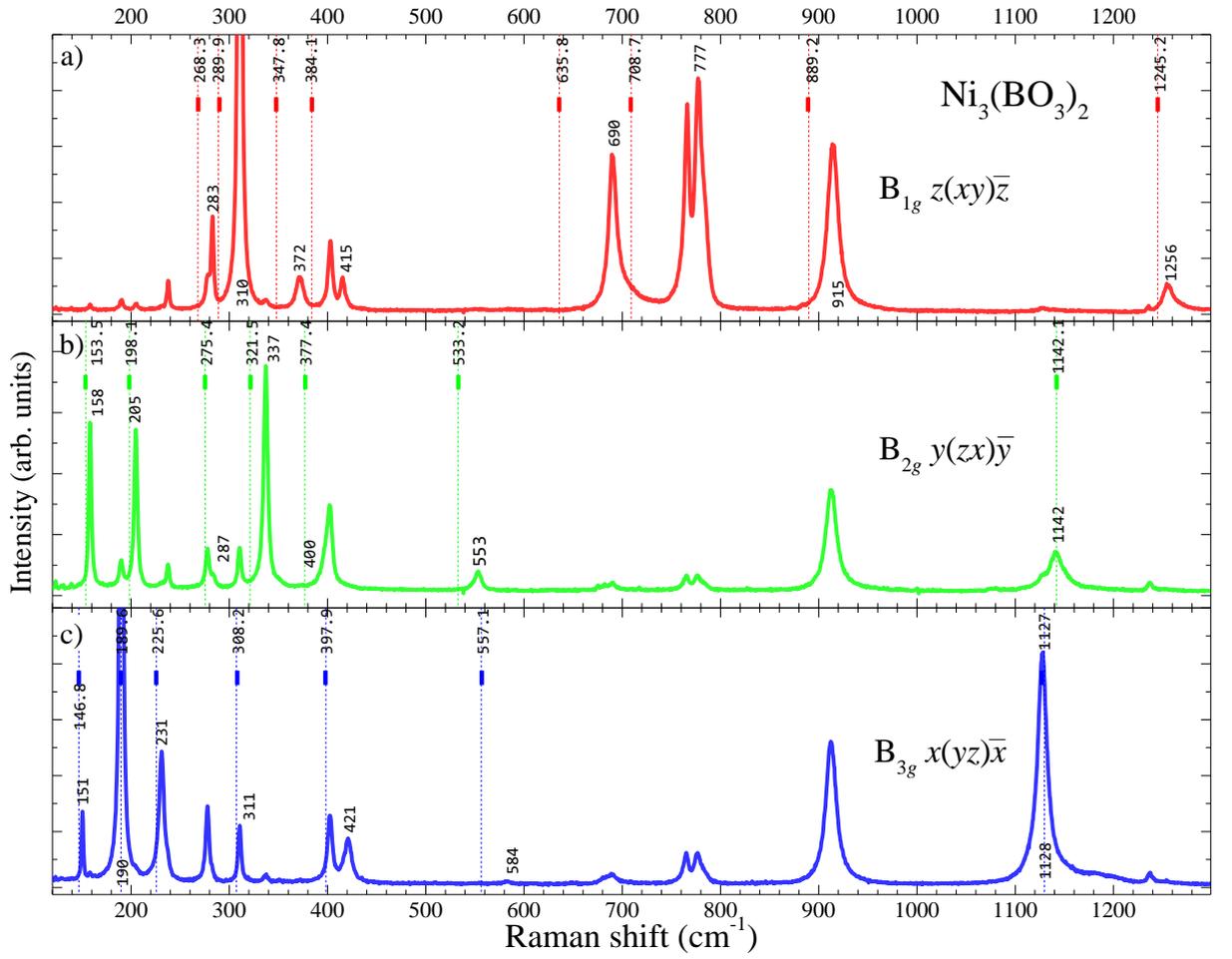

FIG. 7. (Color online) Room-temperature Raman scattering spectra of Ni$_3$(BO$_3$)$_2$ measured in off-diagonal polarizations representing (a) $B_{1g}$ (red line), (b) $B_{2g}$ (green line), and (c) $B_{3g}$ (blue line) modes.

TABLE V. Comparison of calculated and experimental frequencies (cm$^{-1}$) of the $A_g$ phonons.

| DFT | Experiment | Assignment Ni(4f) | [BO$_3$] |
|---|---|---|---|
| 223.20 | 238 | $T_z$ | $T_x$ |
| 264.50 | 278 |  | $T_y$ |
| 326.00 | 351 |  | $R_z$ |
| 377.40 | 403 |  | $T_x$ |
| 638.10 | 681 |  | $v4$ |
| 708.20 | 766 |  | $v2$ |
| **886.30** | **912** |  | **$v1$** |
| 1225.90 | 1238 |  | $v3$ |

TABLE VI. Comparison of calculated and experimental frequencies (cm$^{-1}$) of the $B_{1g}$ phonons.

| DFT | Experiment | Assignment |
|---|---|---|

| | | Ni(4$f$) | [BO$_3$] |
|---|---|---|---|
| 268.30 | 283 | $T_z$ | |
| **289.90** | **310** | | **$T_y$** |
| 347.80 | 372 | | $R_z$ |
| 384.10 | 415 | | $T_x$ |
| 635.80 | 690 | | $\nu 4$ |
| 708.70 | 777 | | $\nu 2$ |
| 889.20 | 915 | | $\nu 1$ |
| 1245.20 | 1256 | | $\nu 3$ |

TABLE VII. Comparison of calculated and experimental frequencies (cm$^{-1}$) of the $B_{2g}$ phonons.

| DFT | Experiment | Assignment | |
|---|---|---|---|
| | | Ni(4$f$) | [BO$_3$] |
| 153.50 | 158 | $T_y$ | |
| 198.10 | 205 | $T_x$ | $T_z$ |
| 275.40 | 287[a,b] | | $T_z$, $R_y$ |
| **321.5** | **337** | **$T_z$** | **$T_z$, $R_y$** |
| 377.40 | 400[a] | | $R_x$ |
| 533.20 | 553 | | $\nu 4$ |
| 1142.10 | 1142 | | $\nu 3$ |

[a] Shoulder-like phonon
[b] Clearly observed at low temperatures

TABLE VIII. Comparison of calculated and experimental frequencies (cm$^{-1}$) of the $B_{3g}$ phonons.

| DFT | Experiment | Assignment | |
|---|---|---|---|
| | | Ni(4$f$) | [BO$_3$] |
| 146.80 | 151 | $T_x$ | $R_y$ |
| **189.60** | **189** | **$T_y$** | **$T_z$** |
| 225.60 | 231 | | $T_z$ |
| 308.2 | 311 | | $R_y$ |
| 397.90 | 421 | | $R_x$ |
| 557.10 | 584[a] | | $\nu 4$ |
| 1127.00 | 1128 | | $\nu 3$ |

[a] Clearly observed at low temperatures

From the Raman scattering results shown above one can deduce the following conclusions. For all the observed Raman lines, counterparts among the computed Raman active modes can be found, although DFT computations systematically underestimate phonon frequencies. Computational results suggest that the doublet spectral structure observed at 766/776 cm$^{-1}$ corresponds to the $A_g$ and $B_{1g}$ modes computed at 708.2 and 708.6 cm$^{-1}$. Similar case is with the broad line near 912 cm$^{-1}$. According to calculations, this line is derived from overlapping of the $A_g$ and $B_{1g}$ phonons that correspond to the Raman Davydov doublet originating from the $v1$ vibration of a free $BO_3$ molecule. These lines are observed in the spectra for both parallel and crossed polarizations, which may point to breaking of the selection rules.

### D. Magnetostructural phase transition at $T_N$ = 46 K and the spin-phonon interaction

Experimental observation of doubling of the magnetic cell [12], competition between FM and AFM exchange interactions and existence of frustrating (disordering) interactions between magnetic $Ni^{2+}$ ions in non-equivalent octahedral positions [10] may serve as a hint of coupling between magnetic and lattice subsystems which may find manifestation in the phonon spectra. Low absorption of $Ni_3(BO_3)_2$ in the FIR region provided us with a possibility to measure the *transmission* spectra by using a sample with a thickness of ~1 mm. This allowed us to observe fine changes in the phonon spectra non detectable in the *reflection* measurements. Figure 8(a) shows transmission spectra of $Ni_3(BO_3)_2$ at temperatures 50 and 40 K just above and below the temperature of the AFM ordering at $T_N$ = 46 K, respectively, measured with the polarization of the incident light $E(\omega)\|x$. The corresponding intensity map in the frequency-temperature axes is presented in Fig. 8(b). Exactly at $T_N$ = 46 K, a striking appearance of narrow weak satellites at 162 and 212 cm$^{-1}$ in a close vicinity of two strong phonon absorption lines is clearly visible. The new lines grow in intensity with further lowering the temperature (see inset in Fig. 8(b)). However, they do not show any noticeable softening or hardening of their positions and therefore cannot be regarded as "*soft*" phonon modes. Both satellite lines at 162 and 212 cm$^{-1}$ are very narrow and have the width of about 1.0 and 0.5 cm$^{-1}$, respectively. The frequency and the temperature behavior of these new modes is in favor of their interpretation as new phonons which appear as a result of a structural phase transition into a less symmetric crystallographic phase. The narrowness of the lines at 162 and 212 cm$^{-1}$ is typical for the so called *folded modes* associated with a folding of the Brillouin zone due to a doubling of the unit cell [66-68]. In this case, lattice phonons from the edge of the BZ fold to the $\Gamma$=0 point and become allowed and observable in the FIR absorption spectra.

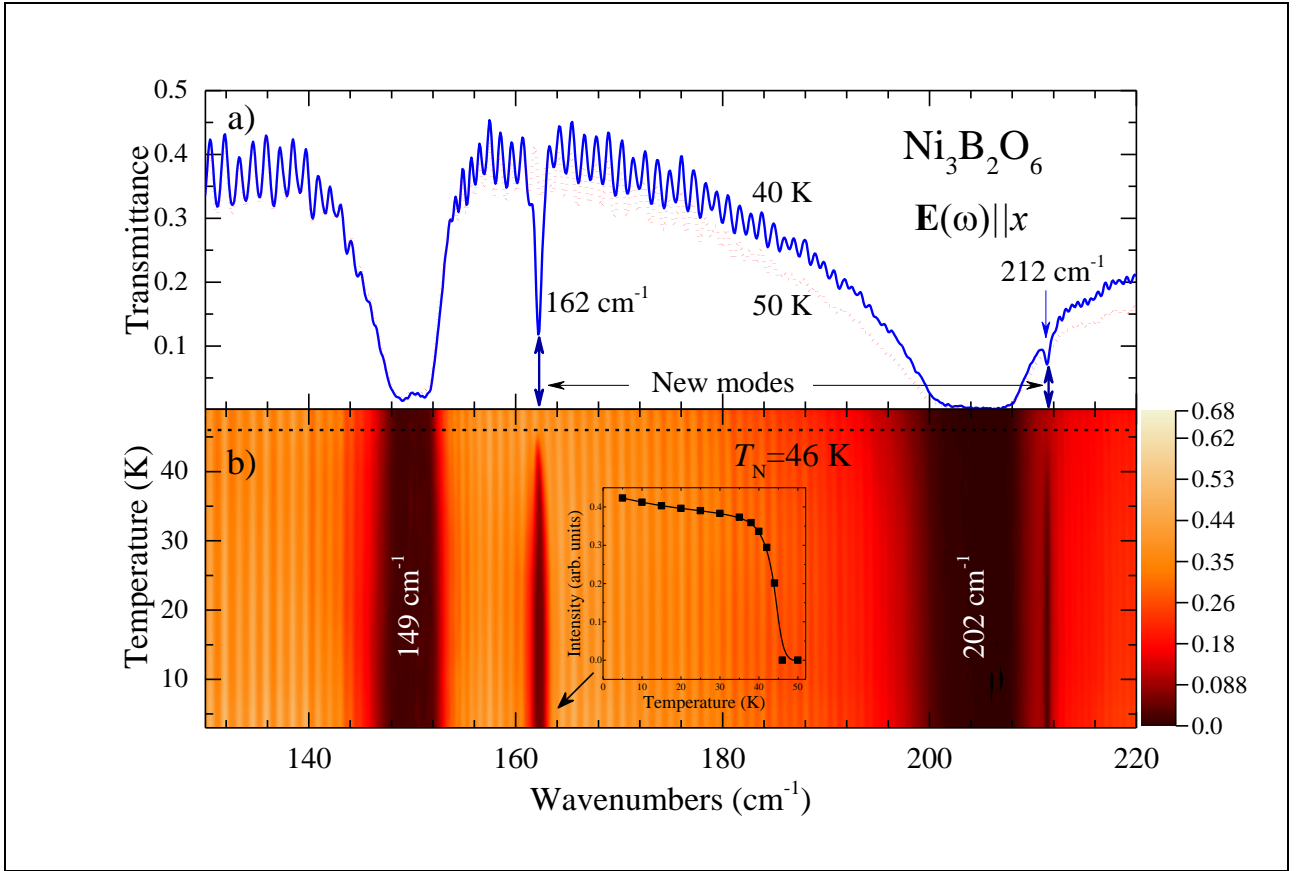

FIG. 8. (Color online). The $E(\omega)\|x$-polarized FIR transmission spectra of $Ni_3(BO_3)_2$ (a) at two temperatures, 50 K > $T_N$ (red dotted trace) and 40 K < $T_N$ (blue solid trace); (b) presented as the reflection intensity map in the frequency-temperature axes. Oscillations in the transmission spectra are due to the interference of incident radiation in a parallel-sided sample. Inset shows the temperature dependence of the intensity of a new phonon line at 162 cm$^{-1}$.

In contrast to the absence of any noticeable softening or hardening of the new folded phonons, some of the "old" phonons demonstrate an additional shift of their frequency in the reflection spectra below the temperature of the magnetic phase transition. Figure 9 shows that, along with a regular hardening of the $B_{3u}$ phonon at 255 cm$^{-1}$ ($T$=250 K) assigned to the $T_x$ translations of the magnetic Ni(2$a$) ions and the [BO$_3$] groups (see Table IV), a noticeable frequency shift of 0.7-1.0 cm$^{-1}$ is observed below $T_N$. Evidently, the mentioned translations modulate the superexchange interaction between the magnetic Ni$^{2+}$ ions (see Fig. 1). This affects elastic constants and, hence, phonon frequencies [39, 66, 69].

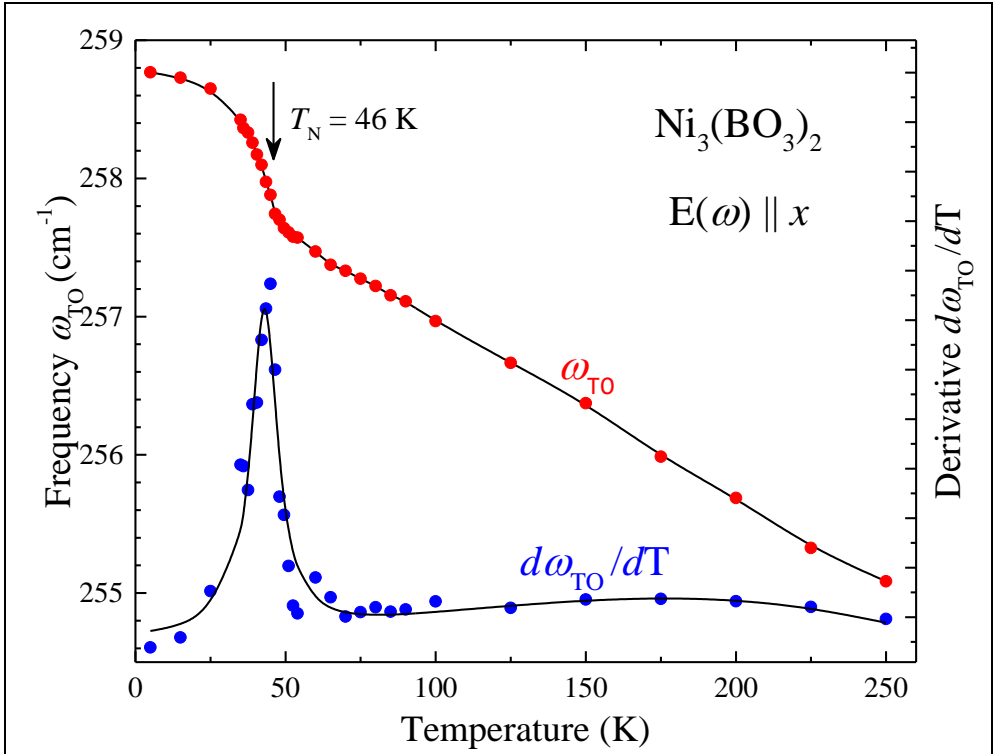

FIG. 9. (Color online). Frequency *vs* temperature dependence of the $B_{3u}$ mode near 255 cm$^{-1}$ (red dots) active in the E($\omega$)||x-polarization and assigned to the $T_x$ translations of the Ni(2a) ions and [BO$_3$] groups. Blue dots show the first derivative $d\omega/dT$ of the phonon frequency, which emphasizes a frequency-shift anomaly at $T_N$. Lines, are guides to the eye.

Next, we focus on the temperature dependence of the $B_{1g}$ Raman-active phonon lines. The results were obtained by fitting spectra to Voigt line shape profiles. Majority of phonon lines show a usual hardening of about 2-4 cm$^{-1}$ (mostly due to anharmonic phonon decay [65]) and line narrowing upon decreasing temperature down to 10 K. However, some of the lines display an anomalous behavior. Figure 10 shows the Raman spectra in the range of 10-50 K, which manifest a strong increase of intensity of forbidden $B_{2g}$ phonons in the $B_{1g}$ spectra below $T_N$, but without any noticeable frequency shift. Important to say, that only $B_{2g}$ phonons show such behavior but no other allowed and forbidden phonons. This observation differs from a usual manifestation of the spin-phonon interaction (shown, for example, in Fig. 9) and strongly suggests a crystallographic symmetry lowering at $T_N$.

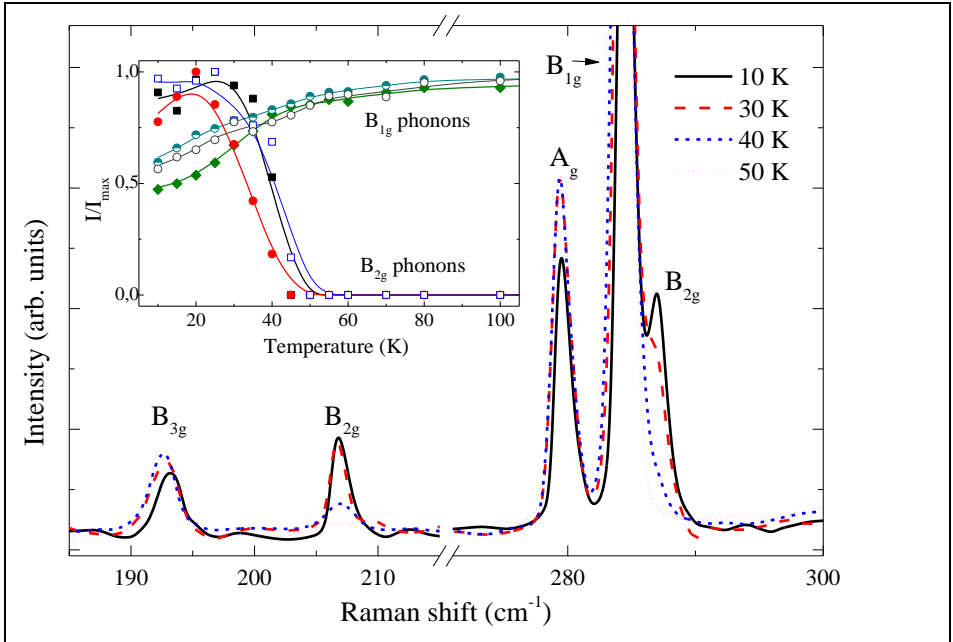

FIG. 10. (Color online) Raman scattering spectra of $Ni_3(BO_3)_2$ for the $z(yx)\bar{z}$ polarization ($B_{1g}$ modes) at temperatures 10, 30, 40, and 50 K. The spectra show a presence of the $A_g$, $B_{3g}$, and $B_{2g}$ forbidden modes. While the $A_g$ and $B_{3g}$ modes diminish in intensity upon cooling, in line with the allowed $B_{1g}$ modes, the $B_{2g}$ ones exhibit a dramatic growth in intensity below $T_N$. Inset: temperature dependences of normalized integral intensities of the phonon lines for the $B_{1g}$ modes with frequencies 283, 310, and 372 cm$^{-1}$ and $B_{2g}$ modes at 205, 287, and 337 cm$^{-1}$. Lines are guides to the eye.

All these observations of temperature dependences of different phonons in the infrared and Raman spectra serve as a clear evidence that the magnetic ordering of $Ni_3(BO_3)_2$ is accompanied by a structural phase transition. Though a magnetic phase transition and a doubling of the magnetic unit cell along the *b* and *c* axes were established in neutron and magnetization measurements [12, 10], no structural transition was noticed. It is worth mentioning, however, that conventional x-ray and neutron diffraction experiments in many cases fail to detect small atomic displacements. For example, it has recently been demonstrated for the rare-earth (*R*) iron oxyborates $RFe_3(BO_3)_4$, using hard x-ray scattering technique [70].

Only a few examples are known of magnetostructural phase transitions in nonmetallic compounds. The most known one is probably the spin-Peierls transition in half-integer spin value AFM chain compounds. Appearance of folded phonon modes was observed at the spin-Peierls transition in $CuGeO_3$ ($Cu^{2+}$ ion, $S=1/2$) [67,71] and at the spin-Peierls-like transition (accompanied by a charge ordering) in α'-$NaV_2O_5$ [68]. Another example is a magnetostructural transition in the geometrically frustrated Heisenberg antiferromagnet on the pyrochlore network $ZnCr_2O_4$, which has been explained in terms of a spin-driven Jahn-Teller effect bearing a resemblance to the spin-Peierls instability in spin chains [72]. Thermally- and light-induced magnetostructural transitions were observed in so-called "breathing crystals" (based on copper-organic-nitroxide exchange-coupled clusters within the polymeric chains) and explained by interplay of exchange interaction between copper spins and the Jahn-Teller nature of copper complexes [73].

The $Ni^{2+}$ ion of $Ni_3(BO_3)_2$ has an integer spin value $S=1$ which rules out the spin-Peierls scenario. A singlet ground state $^3A_2$ of the $Ni^{2+}$ ion in the octahedral crystal field [59,60] eliminates the Jahn-Teller effect as well. All this suggests that we are dealing with a new type of a magnetostructural phase transition which to the best of our knowledge has never been observed before in the nickel nonmetallic compounds. Though the data on $Ni_3(BO_3)_2$ are scarce, in what follows we will qualitatively discuss a possible mechanism of the observed magnetostructural phase transition.

Neutron-diffraction patterns of $Ni_3(^{11}BO_3)_2$ powder samples (specially prepared with the $^{11}B$ isotope to avoid a large neutron capture cross-section of $^{10}B$) taken at 6 K have revealed magnetic reflections that could be indexed in a ($a \times 2b \times 2c$) magnetic unit cell [12]. However, powder pattern intensity data are insufficient for a complete determination of a complex magnetic structure. Recently, magnetization measurements on the $Ni_3(BO_3)_2$ single crystals have shown that, below $T_N=46$ K, magnetic moments of the nickel ion order along the $z(c)$ direction (in our notations, see Sec. II B). It was found that the paramagnetic Curie temperature is very low $\Theta = -7.5$ K and does not depend on the direction of the applied magnetic field [10]. The authors of Ref. [10] have performed a group-theoretical analysis of possible magnetic structures and applied a simple indirect-coupling model to analyze the magnetic structures allowed by symmetry and to estimate the exchange interactions. They have found competing ferromagnetic and antiferromagnetic exchange interactions in the system, which could explain a small value of $\Theta$ as compared to $T_N$. The magnetic structure compatible with both the symmetry predictions and the ($a \times 2b \times 2c$) magnetic unit cell found from neutron scattering powder data [12], as well as with the results of magnetic measurements [10] can be described as follows. The chains of interconnecting triangles formed by nickel ions -Ni(2$a$)-2Ni(4$b$)- are running along the $x(a)$ axis with the magnetic moments of both nickel subsystems being oriented ferromagnetically along the $z(c)$ axis. The neighboring chains are coupled via multiple exchange paths delivering competing ferromagnetic and antiferromagnetic interactions, some of them being frustrated (see Fig. 6 in Ref. [10]). In the magnetically ordered state, a collinear antiferromagnetic ordering of neighboring ferromagnetically ordered chains along the $y(b)$ and $z(c)$ directions takes place, the antiferromagnetic phase possesses the $P_a2_1/c$ magnetic space symmetry group with a doubled primitive cell (2$b$, -$a$, $b+c$, 0, 0, 0) relative to that of the paramagnetic phase $Pnnm$.

On the basis of these findings, we could suggest the following scenario for $Ni_3(BO_3)_2$. Highly frustrated interactions between the ferromagnetic nickel chains prevent a long-range ordering in the system, whereas short-range magnetic fluctuations develop. An interaction of these fluctuations with lattice instability at some point of the BZ triggers a structural phase transition which, in its turn, removes frustrations and a long-range magnetic order is established.

On the assumption of the magnetic space group $P_a2_1/c$ (# 14.80), we find that the corresponding group for atomic positions is $P2_1/c$ (#14). In this case, the number of phonon modes at the BZ center is doubled. The symmetry analysis results in the following irreducible representations and selection rules:

$$132\Gamma = 30A_g(xx,yy,zz,yz) + 36A_u(x) + 30B_g(xy,xz) + 36B_u(y,z) \quad (6)$$

Acoustic modes are $A_u+2B_u$. Correspondence of these low-temperature (LT) modes relevant to phonons below $T_N$ with those of the paramagnetic high-temperature (HT) phase is described by the following relations:

$$30A_g(LT)=8A_g(HT)+ 7B_{3g}(HT)+15M,$$
$$30B_g(LT)=8B_{1g}(HT)+7B_{2g}(HT)+15M,$$
$$36A_u(LT)=7A_u(HT)+ 11B_{3u}(HT)+18M, \quad (7)$$
$$36B_u=7B_{1u}(HT)+11B_{2u}(HT)+18M,$$

where $M$ stands for "the folded" modes from the $(0, ½, ½)$ point of the BZ, and $x, y$, and $z$ refer to the axes of the paramagnetic phase $Pnnm$).

Inspection of the relations (7) shows that (i) IR and Raman active modes from the paramagnetic phase do not mix and remain separated; (ii) IR active LT $A_u$ modes include IR inactive $A_u$ silent modes from the paramagnetic phase; and (iii) Raman active LT $B_g$ modes consist of both $B_{1g}$ and $B_{2g}$ modes of the paramagnetic phase, not counting folded modes. Our experimental data are in agreement with all these conclusions. In particular, appearance of the $B_{2g}$(HT) modes in the (xy) LT Raman spectrum is observed (see Fig. 10). The new 162 and 212 cm$^{-1}$ IR modes (see Fig. 8) could be either formerly silent $A_u$ modes (*ab initio* calculated frequencies are at 161 and 199 cm$^{-1}$), or folded modes. Thus, the phase transition at $T_N$=46 K may be characterized as a magnetostructural transition when both magnetic and structural order parameters are involved. No doubt that further detailed experimental and theoretical studies are necessary to confirm or invalidate the suggested scenario for this intriguing phase transition.

## V. CONCLUSIONS

To summarize, we have carried out a comprehensive study of the lattice phonons at the center of the Brillouin zone for a multi-sublattice antiferromagnet $Ni_3(BO_3)_2$ with a complex orthorhombic crystal structure in which magnetic $Ni^{2+}$ ions occupy two distinct 2(*a*) and 4(*b*) positions. Single crystals were characterized by a spectroscopic ellipsometry technique in the range of 0.6-5.6 eV and showed substantial changes of the dielectric functions in comparison to the simplest nickel oxide NiO. Experimental studies were performed using infrared reflection and transmission, as well as Raman scattering spectroscopy. The obtained results were supported by theoretical *ab initio* DFT calculations. All the odd and even phonons predicted by the symmetry analysis were found experimentally and their one-to-one correspondence to the calculated modes was established. These results have delivered knowledge on eigenvectors of particular observed lattice vibrations. Appearance of several new phonon modes and anomalous behavior of some of the "old" phonons at the antiferromagnetic ordering temperature $T_N$=46 K and below delivered a clear prove of a structural phase transition intimately related to magnetic ordering of $Ni_3(BO_3)_2$. This structural transition was not noticed in previous studies. Obviously, it is associated with interaction between magnetic and lattice subsystems in $Ni_3(BO_3)_2$. Coupling of magnetic fluctuations in quasi-one-dimensional ferromagnetic chains formed by Ni(2*a*)-2Ni(4*b*) triangles, interconnected by highly frustrated antiferromagnetic interactions, to an unstable phonon branch could be a driving force of this magnetostructural phase transition. This force removes frustrations and, thus, promotes a 3D magnetic ordering. In addition, a clear evidence of the spin-phonon interaction was observed below $T_N$ for particular phonons and these results support the scenario of a complicated coupling between the lattice and spin dynamics in $Ni_3(BO_3)_2$. We believe that our observations and conclusions will stimulate further theoretical and experimental studies of not only $Ni_3(BO_3)_2$ but also other materials with the kotoite structure.

**ACKNOWLEDGEMENTS**

We thank G. T. Andreeva for growing single crystals of $Ni_3(BO_3)_2$, and N. F. Kartenko and A. S. Kolosova for the x-ray orientation of the samples. We thank Yu. G. Pashkevich for fruitful discussions of the magnetic structure of $Ni_3(BO_3)_2$. This work was supported by the Russian Government under the Project No. 14.B25.31.0025, by the Russian Foundation for Basic Research (Projects Nos. 15-02-04222a and 15-32-20613), and by the Russian Science Foundation (Project No. 14-12-01033). Ellipsometry measurements were performed at the Radboud University Nijmegen under the INTAS Project 1000008-7833. The computations were performed using facilities of the Computational Centre of the Research Park of St. Petersburg State University.